\documentclass[aps,pre,twocolumn,superscriptaddress]{revtex4}

\usepackage{pb-diagram}
\usepackage{fancyhdr}
\usepackage{CJK}
\usepackage{graphicx}
\usepackage{appendix}
\usepackage{listings}
\usepackage{url}
\usepackage{lineno,hyperref}
\usepackage{color}
\usepackage{amsmath,amsthm,amssymb,amsfonts}
\usepackage{mathrsfs}
\usepackage{tabularx}
\usepackage{bm,braket}

\def\d{{\rm d}}
\def\tr{{\rm tr}}
\renewcommand{\eqref}[1]{\textrm{Eq.}(\ref{#1})}

\newcolumntype{Y}{>{\centering\arraybackslash}X}

\begin{document}


\begin{CJK*}{UTF8}{gbsn}
\title{Ergodicity and Mixing in Quantum Dynamics}
\author{Dongliang Zhang(张东良)}
\affiliation{School of Physics, Peking University, Beijing 100871, China}

\author{H. T. Quan (全海涛)}
\affiliation{School of Physics, Peking University, Beijing 100871, China} 
\affiliation{Collaborative Innovation Center of Quantum Matter, Beijing 100871, China}

\author{Biao Wu(吴飙)}
\email{wubiao@pku.edu.cn}
\affiliation{International Center for Quantum Materials, School of Physics, Peking University, Beijing 100871, China} 
\affiliation{Collaborative Innovation Center of Quantum Matter, Beijing 100871, China}
\affiliation{Wilczek Quantum Center, College of Science, Zhejiang University of Technology,  Hangzhou 310014, China}





\begin{abstract}
After a brief historical review of ergodicity and  mixing in dynamics, 
particularly in quantum dynamics,  we introduce definitions of 
quantum ergodicity and mixing using  the structure of the system's energy 
levels and spacings. Our definitions are consistent with usual understanding 
of ergodicity and mixing. Two parameters concerning the degeneracy 
in energy levels and spacings are introduced.  
They are computed for right triangular billiards and the results indicate 
a very close relation between quantum ergodicity (mixing) and quantum chaos.  
At the end, we argue that, besides ergodicity and mixing,  there may exist a 
third class of quantum dynamics which is characterized by a 
maximized entropy.
\end{abstract}


\maketitle
\end{CJK*}
\section{Introduction}\label{sec:intro}
Ergodicity and mixing are of fundamental importance in statistical 
mechanics. Ergodicity justifies the use of microcanonical ensemble and 
mixing ensures that a system approach equilibrium dynamically~\cite{Penrose}. 
However,  it is difficult to prove with mathematical rigor that a classical dynamical 
system  is ergodic or mixing.  As a result,  the microcanonical ensemble in textbooks 
is still established with postulates~\cite{Huang}.  More importantly, the concept of 
ergodicity and mixing is now obsolete in the following sense:  they are defined 
for classical dynamics while the dynamics of microscopic particles are 
fundamentally quantum. To establish statistical mechanics with quantum dynamics,  
we need to define ergodicity and mixing in quantum dynamics.  

In this work we define quantum ergodicity and mixing using the structure 
of the system's energy levels and spacings without any assumption.
With  the early results by von Neumann~\cite{neumann1929,neumann1929beweis} 
and Reimann~\cite{reimann2008}, it can be shown that our 
definitions, which appear very mathematical, do lead to the usual
understanding of ergodicity and mixing. We introduce two parameters to 
characterize the degeneracy in energy levels and spacings, respectively. 
They are computed numerically for right triangular billiards, whose 
classical dynamical properties have been studied in great 
detail~\cite{casati1997,wang2014}.  The numerical results indicate
that there is a very close relation between quantum chaos and quantum 
ergodic or mixing:  most of non-integrable finite quantum systems are 
both ergodic and mixing.  It is clear from this example that a system 
whose quantum dynamics is ergodic may not be ergodic in the 
corresponding classical dynamics.

We draw a parallel between our paper and Peres's two papers~\cite{peres1984,peres2}.
Peres introduced his definitions for quantum ergodicity and mixing in the
first paper~\cite{peres1984} then illustrated with 
examples  these two concepts with his co-authors in the second paper~\cite{peres2}. 
In our paper we do both: we first introduce our definitions for quantum ergodicity and 
mixing and then illustrate them with examples. 


\section{Quantum Ergodicity and Mixing\label{sec:def}}
\subsection{History}
Ergodicity was introduced by Boltzmann in 1871 as a hypothesis to understand 
thermodynamics microscopically~\cite{Penrose}. Mixing was first discussed by Gibbs~\cite{GibbsBook} and its mathematical definition was introduced by von Neumann 
in 1932~\cite{Neumann1932}. Both concepts concern the long time behavior 
of dynamical systems and are of fundamental importance to statistical mechanics. 
They are now the focus of a fully-developed 
branch of mathematics called ergodic theory\cite{EH2,EH}.  However, ergodicity and mixing 
are becoming less interesting to physicists for two reasons: (1) After decades of research 
with many meaningful results\cite{EH2,EH}, it is still not rigorously proved that 
many of the physical systems existing in nature  are either ergodic or mixing. 
(2) Both ergodicity and mixing are 
only defined for classical systems while the microscopic particles are fundamentally quantum.  
Therefore, it is imperative to define both ergodicity and mixing to quantum systems. 

The first physicist who discussed quantum ergodicity was von Neumann. In 1929, 
von Neumann proved two inequalities, which he named quantum ergodic theorem and 
quantum $H$-theorem~\cite{neumann1929beweis}, respectively. His ergodic theorem 
ensures that not only the long 
time average of a macroscopic observable equals to its microcanonical ensemble average 
but also has small fluctuations. In other words,  the observable deviates considerably from 
its averaged value only rarely.  So, by ergodicity von Neumann meant actually both ergodicity 
and mixing.  Interestingly, mixing as defined for classical dynamics was only introduced 
three years later in 1932 by von Neumann~\cite{Neumann1932}. In addition, 
according to von Neumann's $H$-theorem, once the quantum system dynamically relaxes
to its equilibrium state, where macroscopic observables have small fluctuations, this state
also has a maximized entropy. 

Von Neumann's results had been criticized by many~\cite{peres1984,goldstein}. 
We share the view by Goldstein {\it et al.}~\cite{goldstein} that 
the criticism was mostly misguided; von Neumann had captured
the essence of quantum ergodicity and mixing and his results are inspirational. 
Nevertheless, there do exit some issues with von Neumann's results. 
Most of the variables involved in the two theorems are not computable 
in principle~\cite{han2014}. The reason is as follows. 
To prove his theorems, von Neumann introduced a coarse-graining, which groups the 
Planck cells in quantum phase space into some big cells. All the microscopic 
states in one group of Planck cells correspond to a single macroscopic state. This kind of 
coarse-graining is certainly reasonable. However, no one knows how to technically 
establish such many-to-one mapping between macroscopic state and microscopic state. 
This makes many of von Neumann's variables in Ref.~\cite{neumann1929beweis} uncomputable. Though there may be some revisions to the theorem~\cite{asadi2015quantum}, this difficulty 
is not overcome. 

More recent definitions of ergodicity and mixing in quantum mechanics were 
given by Peres~\cite{peres1984}. Peres recalled the behavior of dynamical 
variables in classical ergodic and mixing systems and expected that there should 
be analogous behavior in quantum ergodic and mixing systems. In accordance 
with von Neumann's results, Peres defined ergodicity as the time average 
of any quantum operator equal to its microcanonical ensemble average and  mixing as 
any quantum operator having small fluctuations.  However, Peres's definitions were 
based on his own definition of quantum chaos~\cite{peres1984},
which is a subject of debate itself. To define quantum chaos, 
Peres  used an ambiguous concept of pseudorandom matrix.  These two steps that
are not mathematically very rigorous , along
with other reasonable but ambiguous assumptions, render Peres's definitions
not satisfactory.


In literature quantum ergodicity has been studied from a different perspective, where
the concept ``quantum ergodicity" is regarded as a branch of quantum chaos  \cite{shnirel1974,feingold1986,zelditch1987,Chirikov1988,zelditch1996billiards,zelditch1996quantum,Casati1996,backer1998,zelditch2005,barnett2006}.  This group of researchers mainly focused on how the eigenfunctions of a Hamiltonian converges to equidistribution in classical phase space in the semi-classical limit or high energy limit with little discussion on dynamical behavior of a quantum system. Their definitions of quantum ergodicity and mixing rely on the corresponding classical cases and thus are not genuinely quantum mechanical. 
Srednicki's work on eigenfunction thermalization follows along this line and has
little discussion on quantum dynamics~\cite{Srednicki1994}. 

In the following subsection we define quantum ergodicity and mixing 
using the energy structure of the system, that is, eigen-energies and their 
spacings. Our definitions are mathematically precise and have no assumptions. 
Furthermore,  by following von Neumann~\cite{neumann1929} and 
Reimann~\cite{reimann2008},  we can show that our definitions lead to the usual 
physical understanding of ergodicity and mixing. 
Our definitions are based on quantum dynamics and expressed in the language of 
pure quantum mechanics without referring to classical mechanics.  
Near the end of this paper, based on a recent work~\cite{han2014}, we argue that we may 
be able to expand our definitions
to include a third class of quantum dynamics, which is charactering by a maximized 
entropy. 



\subsection{Definitions}
\label{sec:def2}
Consider a quantum system with discrete eigen-energies $\{ E_n \}$ and 
corresponding energy eigenstates $\{ | \phi_n \rangle \}$. 
\theoremstyle{definition} \newtheorem*{ergodicity}{Ergodicity}
\begin{ergodicity}
A quantum system is ergodic if its eigen-energies satisfy
\begin{equation}\label{nde}
\delta_{E_m,E_n}=\delta_{m,n}\,.\tag{I}
\end{equation}
\end{ergodicity}
\theoremstyle{definition} \newtheorem*{mixing}{Mixing}
\begin{mixing}
A quantum system is mixing if its eigen-energies satisfy
both the above condition (I) and the following condition
\begin{equation}\label{ndg}
\delta_{E_k-E_l,E_m-E_n}=\delta_{k,m}\delta_{l,n}\,,~~~ \text{for }   k\neq l, m \neq n\,.\tag{II}
\end{equation}
\end{mixing}
Condition (I) indicates that there is no degenerate eigenstate. Condition (II) implies 
that there is no degeneracy in energy gaps between any pair of eigen-energies.  
It is clear that a quantum system that is mixing must be ergodic, similar to classical 
dynamics.  As we shall show in the following,  condition (I) can lead to the usual intuitive 
understanding of ergodicity: the long time average equals to the ensemble average. 
With both conditions (I) and (II),  one can show that the so-defined mixing indeed 
means a small time fluctuation for an observable. 

Suppose that the quantum
system is in an initial state $| \psi (0) \rangle =\sum\nolimits_n c_n| \phi_n \rangle$. 
After evolving for a period of time $t$, the quantum system is in a state described by 
$| \psi (t) \rangle =\sum\nolimits_n c_ne^{-iE_n t/\hbar}| \phi_n \rangle$.  
For an observable $\hat{A}$, 
its expectation value at time $t$ is given by
\begin{equation}
\langle \hat{A}(t)\rangle \equiv\langle \psi (t)  | \hat{A} | \psi (t) \rangle = \tr \hat{A}\hat{\rho} (t),
\end{equation}
where $\hat{\rho}(t) \equiv | \psi (t) \rangle\langle \psi (t) |$ is the density matrix at time $t$. 
Its long time average is
\begin{equation}
\langle \hat{A}\rangle_T \equiv \lim_{T\rightarrow \infty}\frac1T \int_0^T
\langle \hat{A}(\tau)\rangle d \tau\,.
\end{equation}
We now introduce a density matrix
\begin{equation}\label{rhomc}
\hat{\rho}_{\text{mc}} \equiv \sum_n |c_n|^2|\phi_n \rangle \langle \phi_n |\,.
\end{equation}
This density matrix does not change with time and it can be regarded as 
describing a micro-canonical ensemble~\cite{neumann1929,reimann2008,han2014}. 
This allows us to define the ensemble average as 
\begin{equation}
\langle \hat{A} \rangle_E\equiv \tr \hat{A}\hat{\rho}_{\text{mc}}\,.
\end{equation}
For a quantum system satisfying condition (I),  it is easy to 
check that~\cite{neumann1929,peres1984} (see also Appendix B)
\begin{equation}
\label{ergodic}
\langle \hat{A}\rangle_E=\langle \hat{A}\rangle_T\,, 
\end{equation}
that is, the long time average of $\hat{A}$ equals to its micro-canonical 
ensemble average.

For a quantum system satisfying both conditions (I) and (II), one can 
prove~\cite{reimann2008,short2011} that  the long-time  averaged
fluctuation $F_A^2$ satisfies (see also Appendix B)
\begin{equation}\label{mixing}
F_A^2 \equiv 
\frac{\left\langle \Big| \langle \hat{A}(t)\rangle- \langle \hat{A}\rangle_E\Big|^2 \right\rangle_T}{\|\hat{A}\|^2}
 \leq \tr \hat{\rho}^2_{\text{mc}},
\end{equation}
where $\| \hat{A} \|^2=\sup \{\langle \psi | \hat{A}^\dag \hat{A} | \psi \rangle : \ket{\psi} \in \mathscr{H}\}$ is the upper limit of
the expectation value of $\hat{A}^2$ in the Hilbert space.  This demonstrates that
a mixing quantum system indeed has small time fluctuations. 

\theoremstyle{remark} \newtheorem{R}{Remark}
A few remarks are warranted here to put our definitions in perspective. 
\begin{R}
Our definitions of ergodicity and mixing for quantum systems are mathematically very precise . 
They do not involve any concepts and assumptions, which are mathematically ambiguous. 
Peres made many assumptions in his definitions~\cite{peres1984},
which are reasonable but ambiguous mathematically. In particular, we do not need to define
quantum chaos first as Peres did~\cite{peres1984}. 
\end{R}
\begin{R}
Although our definitions appear very mathematical,  as we have shown, they 
are consistent with the familiar physical pictures that we have had
with ergodicity and mixing in classical dynamics: 
ergodicity means that long time average equals to 
ensemble average; mixing implies small time fluctuations.    Moreover, 
similar to the classical case, a quantum mixing system is ergodic but not vice versa. 
\end{R}
\begin{R}
Our definitions have their roots in the 1929 paper, where von Neumann 
proved a quantum ergodic theorem~\cite{neumann1929}. However, 
von Neumann in 1929 did not distinguish between ergodicity and mixing. 
His view of ergodicity at that time is closer to the current view of mixing. 
In other words, his quantum ergodic theorem may be better called quantum 
mixing theorem. 

It is worthwhile to note two interesting points: (i) von Neumann used both condition (I) and (II) to 
prove his quantum ergodic theorem; (ii) mixing in classical dynamics
was introduced three years later in 1932 by von Neumann himself~\cite{Neumann1932}. 
\end{R}
\begin{R}
The density matrix $\hat{\rho}_{\text{mc}}$ is used as the micro-canonical ensemble in the above
discussion. It is not the standard micro-canonical ensemble found in textbooks~\cite{Huang},
\begin{equation}
\hat{\rho}_{\text{tb}}=\frac1N \sum_{E_n\in[E, E+\delta E] }|\phi_n\rangle\langle\phi_n|\,,
\end{equation}
where $N$ is the number of energy-eigenstates in energy interval $[E, E+\delta E]$.
However, we can certainly choose an initial state such that $|c_n|^2=1/N$ for 
$E_n\in[E, E+\delta E]$ and $|c_n|=0$ otherwise. In this way, we recover the textbook
micro-canonical ensemble $\hat{\rho}_{\text{tb}}$. That is, $\hat{\rho}_{\text{tb}}$
is just a special case of $\hat{\rho}_{\text{mc}}$.  
\end{R}
\begin{R}
Our definitions of ergodicity and mixing for quantum systems  are independent of 
initial conditions. Nevertheless, to thoroughly understand them, we do need to consider
initial conditions as the density matrix $\hat{\rho}_{\text{mc}}$ depends on initial conditions. 
While \eqref{ergodic} and \eqref{mixing} hold for an arbitrary initial condition, 
not all of the initial conditions are of physical interest. 
If we choose an initial state where only a few eigenstates are occupied, not only 
$\hat{\rho}_{\text{mc}}$ is no longer sensible to be regarded as a micro-canonical ensemble
but also the fluctuation $F_A^2 $ in Eq.(\ref{mixing}) is not small. 
However, this kind of initial conditions are hard to realize in experiment or to be found in nature for a many-body quantum system. Physically, when a many-body quantum system is excited by a practical means, it usually enters into a quantum state where a large number of eigenstates are occupied. 
This is also the reason that  the standard micro-canonical ensemble 
$\hat{\rho}_{\text{tb}}$, which looks quite artificial, works well as long as the system is large. 

This aspect is quite similar to classical systems. In an ergodic or mixing classical system 
there always exist solutions which are not ergodic or mixing, for example, the periodic orbits. 
However, these non-ergodic or non-mixing solutions are rare or have measure zero 
in rigorous mathematical language so that the overall properties of the system are not affected.
\end{R}

\begin{R}
Many quantum systems have certain symmetries and correspondingly some 
good quantum numbers. Energy degeneracy can easily occur between 
the eigenstates of different symmetric sectors. As a result, 
these quantum systems are in general not ergodic and mixing. However, if one
focuses only one symmetric sector, the quantum system can be ergodic or mixing. 
In this case, we may say that the quantum system is ergodic or mixing 
in a sub-Hilbert space. 
\end{R}

\begin{R}
Although classical ergodicity and mixing are of fundamental importance in 
statistical mechanics,  their definitions can be applied to single-particle systems. 
Similarly, our definitions can be applied to single-particle quantum systems. 
\end{R}

To conclude our definitions we offer two simple and illustrative examples.
The first is one dimensional harmonic oscillator. 
There is no energy degeneracy so it is ergodic. There is a great deal of degeneracy
in energy spacings so that it is not mixing. Interestingly, for the classical dynamics, 
the one dimensional harmonic oscillator is similar: it is ergodic but not mixing. 

The second example is a particle in a one dimensional box system, where there is neither degeneracy in energy levels nor in energy spacings. According to our definitions, it is both quantum ergodic and mixing. It is not difficult to find that its classical counterpart is indeed both ergodic and mixing~\cite{MaBook}.

\subsection{Degeneracy parameters}
There has been a tremendous amount work on quantum chaos or 
quantum non-integrability~\cite{GutzwillerBook}. It is 
interesting to see how our quantum ergodicity 
or mixing is related to quantum chaos. In other words, are the two
conditions (\ref{nde}) and (\ref{ndg}) easy to satisfy in quantum chaotic 
systems? 

On the other hand, it is also interesting to know how infrequent exceptions to these two 
conditions affect quantum dynamics. When  a quantum system  has a relatively small number of  
degeneracies, then almost all its states contain either no degenerate eigenstates or only a few. 
For the former, Eq.(\ref{ergodic}) still holds; for the latter,  the left hand side in 
Eq.(\ref{mixing}) differs the right hand side only slightly. So, in a practical sense, 
this quantum system is ergodic.  The situation is similar for mixing: infrequent degeneracy 
in energy gap is not important. Von Neumann had a similar point of view~\cite{neumann1929}.  
Short and Farrelly showed quantitatively how infrequent degeneracy are not 
important~\cite{Short2012}. 

To address the above two issues, 
we introduce two parameters $\zeta$ and $\xi$,  which describe the 
average degeneracy in energy levels and average degeneracy in energy level spacings, 
respectively, for a given finite set of  energy levels.  The parameter  $\zeta$ is defined as 
\begin{equation}
\zeta = \frac1{N}{\sum\limits_{m,n}\left(\delta_{E_m,E_n}-\delta_{m,n}\right)}\,,
\end{equation}
where $N$ is the number of energy levels in the set. The other parameter  $\xi$ 
is defined as 
\begin{equation}
\xi = \frac{1}{N(N-1)}\sum\limits_{k\neq l,m\neq n}
(\delta_{E_k-E_l,E_m-E_n}-\delta_{k,m}\delta_{l,n})\,.
\end{equation}

Furthermore, it is useful to define two distribution functions, $f(\epsilon)$ and $g(\Delta)$. 
$f(\epsilon)$ is the probability of the eigen-energies having value $\epsilon$; $g(\Delta)$ is
the probability of the energy level spacings at $\Delta$.  With the aid of these distribution 
functions, we can reformulate the definitions of $\zeta$ and $\xi$, respectively, as 
\begin{equation}\label{defzeta}
\zeta=N\sum\limits_\epsilon f^2(\epsilon)-1\,,
\end{equation}
\begin{equation}\label{defxi}
\xi=N(N-1)\sum\limits_\Delta g^2(\Delta)-1\,.
\end{equation}
These two functions are clearly related; their explicit relation is  
\begin{eqnarray}
g(\Delta)=&&\sum_\epsilon f(\epsilon)\frac{Nf(\epsilon+\Delta)-\delta_{\Delta ,0}}{N-1}\nonumber\\
=&&\frac{N}{N-1}\sum_\epsilon f(\epsilon)f(\epsilon+\Delta)-\frac{\delta_{\Delta ,0}}{N-1}.
\end{eqnarray}
We clarify that in our definition $\Delta$ can be negative. In other words, for two arbitrary energy levels $E_m$ and $E_n$ with $m \neq n$, $E_m-E_n$ and $E_n-E_m$ give rise to two energy level spacings rather than one.  We also emphasize that our definitions of $\zeta$ and $\xi$ are for a given set of
energy levels not for all the energy levels in the system. The reason is that only a finite set of energy levels
are involved in any meaningful physical process. 

According to these definitions, the two conditions (\ref{nde}) and (\ref{ndg}) are equivalent 
to $\zeta \equiv 0$ and $\xi \equiv 0$, respectively. The larger $\zeta$(or $\xi$) is  the stronger the non-degenerate-energy condition (\ref{nde}) (or the non-degenerate-gap condition (\ref{ndg})) is violated. We anticipate that for small $\zeta$ and $\xi$ quantum systems can still
be regarded as  ergodic or mixing.  For systems where $\zeta$ and $\xi$ are strictly equal to zero,
we say that they are {\it ideal} ergodic systems or {\it ideal} mixing systems. 

\section{Model}\label{sec:model}
\begin{figure}
\includegraphics[width=0.8\linewidth]{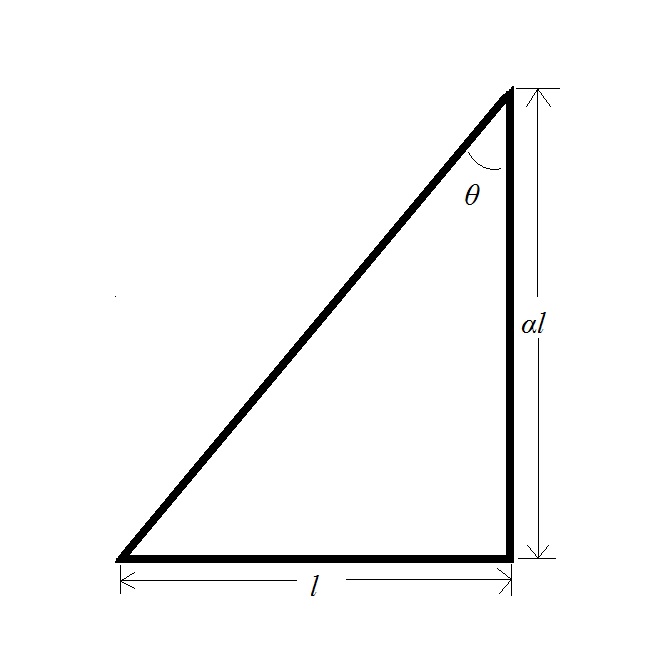}
\caption{A right triangle billiard. Without loss of generality, we take  $\alpha \geq 1$ or, equivalently, $0< \theta \leq \pi/4$. }
\label{Tbilliard}
\end{figure}

In this section we use an example to illustrate our concepts of quantum ergodicity and mixing. 
We consider the motion of one particle with mass $m$ in a right triangular billiard, as is shown in Fig.\ref{Tbilliard}. Mathematically, this billiard is described by the following potential
\begin{equation}
V(x,y)=
\left\{
\begin{aligned}
&0 &0<x<l, 0<y<\alpha x\\
& \infty& \text{otherwise}
\end{aligned}
\right.\,.
\end{equation}
Without loss of generality, we restrict ourselves to  $\alpha \geq 1$ or
$0< \theta \leq \pi/4$ ($\alpha=\cot{\theta}$). It is interesting to note that 
this billiard system is equivalent to the system of  two hard-core particles moving in 
one-dimensional square potential with infinite walls~\cite{wang2014}.  

We choose  this simple  model for two reasons. (1) We can study both 
the integrable cases and chaotic cases by adjusting $\alpha$.  
(2) Many meaningful results on  classical ergodicity and mixing 
in this model have been obtained previously~\cite{casati1997,wang2014}, 
and we can compare them to our quantum results. 
Other models  such as the Bose-Hubbard model~\cite{Benet2010} do not have the above advantages.

The classical integrability of this model is well known. 
The system is integrable only when $\theta=\pi /4$ or $\theta=\pi /6$ (equivalently,  $\alpha=1$ 
or $\alpha=\sqrt{3}$).  
When  $\theta=\pi M/N$, where $M$ and $N$ are two coprime integers and $(M,N) \neq (1,4),(1,6)$, the system has two independent invariants. However, it is regarded as  pseudointegrable\cite{berry1981,gorin2001,zyczkowski1992} because the invariant surface of classical motion in phase space has a genus $2\leq g < \infty$ (it is integrable only when the genus $g=1$). For all other values of $\theta$, the triangle system 
has only one invariant and is generally regarded as chaotic. 

To study the quantum dynamics of  this model, we need to calculate the eigenenergies and 
eigenstates. This can be done only numerically for an arbitrary value of parameter $\alpha$. 
We use the exact diagonalization method (see Appendix \ref{app} for details). 
In our calculation, we choose  $h=m= l=1$. In addition, to avoid confusion, we use 
the single parameter $\alpha$ (instead of $\theta$) to represent the shape of the 
triangle billiard in following discussion.


\section{ Distributions of Energy Levels} \label{sec:eng}
In Section \ref{sec:def} we have defined quantum ergodicity 
and mixing with two conditions (I) and (II) that regard 
the distribution of the system's energy levels. In this section, 
we shall examine to what extent these two conditions are 
satisfied by the triangle billiard and how they are related
to the integrability of this model via parameters $\zeta$
and $\xi$.   In the next section, we shall show that 
the quantum dynamics of the triangle billiard are dictated by
these two conditions. 

\begin{figure*}[ht]
\begin{minipage}{0.39\linewidth}
\centering
\includegraphics[width=\linewidth]{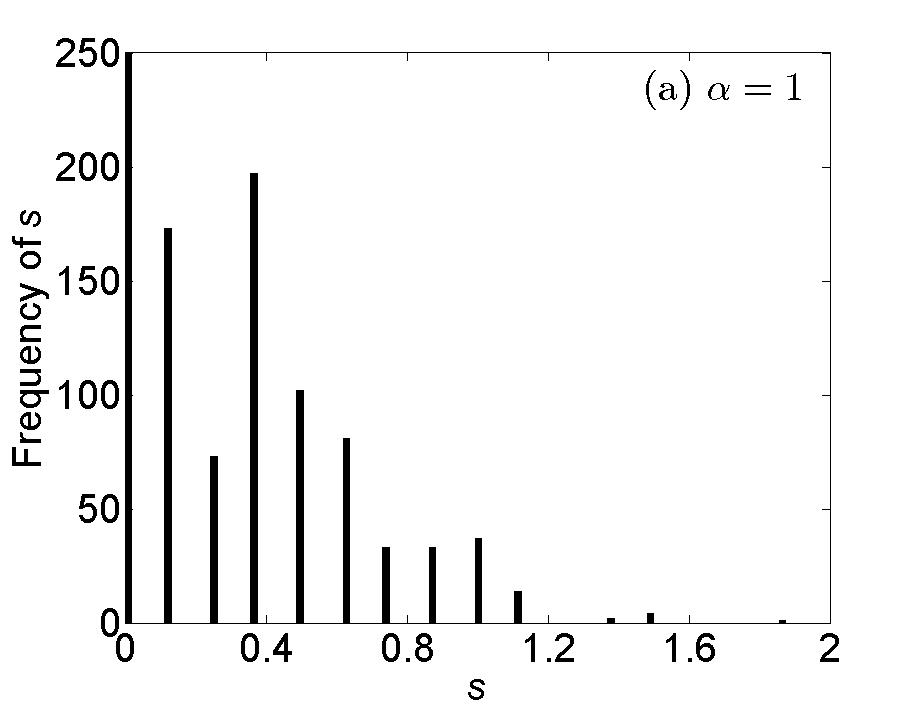}
\end{minipage}
\begin{minipage}{0.39\linewidth}
\centering
\includegraphics[width=\linewidth]{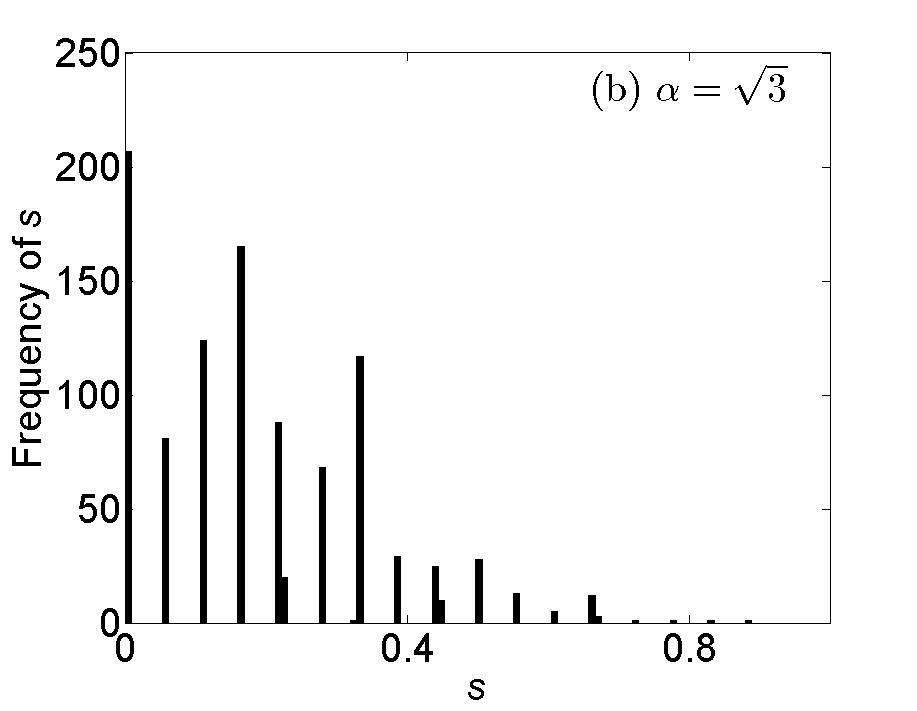}
\end{minipage}
\begin{minipage}{0.39\linewidth}
\centering
\includegraphics[width=\linewidth]{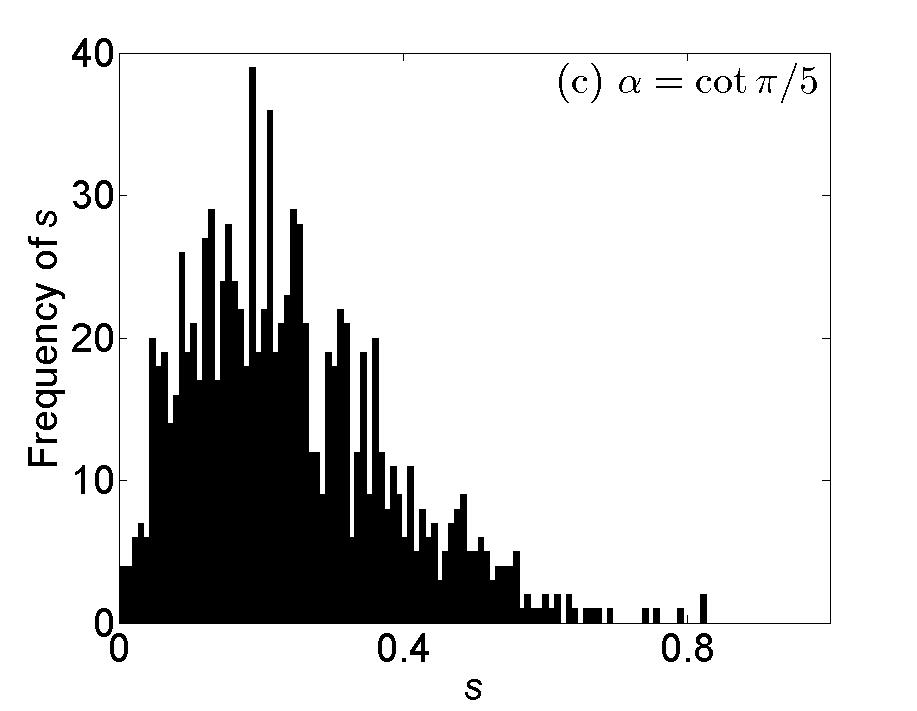}
\end{minipage}
\begin{minipage}{0.39\linewidth}
\centering
\includegraphics[width=\linewidth]{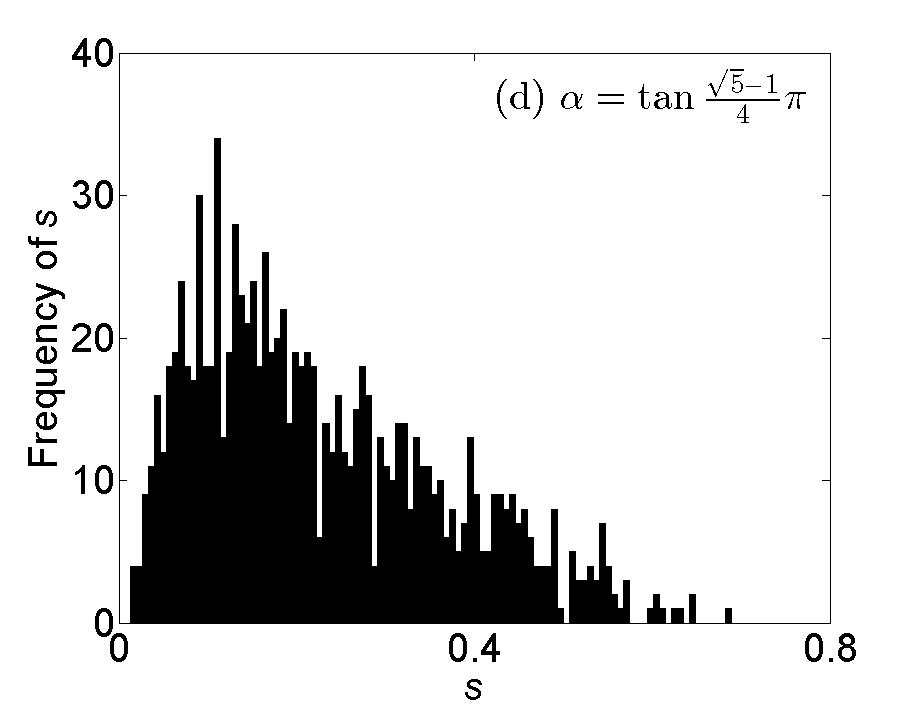}
\end{minipage}
\caption{Nearest spacing distribution of eigenenergies. (a)$\alpha=1$ and (b)$\alpha = \sqrt3$ are two integrable cases. 
(c) $\alpha = \cot(\pi/5)$ is pseudointegrable. (d) is chaotic.  Calculations are done in first 1000 energy levels. }
\label{nsd}
\end{figure*}

The study of quantum chaos has revealed that the structure of quantum energy levels 
of a system is closely related to  the classical integrability of the system~\cite{peres1984,reichl2004}. 
One often uses  the nearest spacing distribution(NSD) $p(s)$ of a system to describe 
its structure of quantum energy levels, where $s>0$
is the spacing between two nearest energy levels. The following feature 
of NSD  is well known.  For a system whose classical dynamics is integrable, 
its NSD is Poisson-like with a peak distribution at $s=0$.  For a system whose 
classical dynamics is chaotic,  the NSD of its quantum energy levels
is Wigner-like: an almost zero probability density at $s=0$ and a peak 
density at $s=s_m \not= 0$. This feature indicates that condition
(I) is always satisfied by a quantum chaotic system. 
There is no clear conclusion for condition (II) as the peak at nonzero 
$s$ in the Wigner distribution seems to suggest that condition (II) 
is not satisfied by a quantum chaotic system.  However, our following
numerical results show that condition (II) is also largely satisfied by a quantum chaotic system.


The Hamiltonian matrix of  the triangle billiard is diagonalized 
numerically for a set of $\alpha$. Its NSDs for the first 1000 
energy levels are shown in Fig.\ref{nsd} for four typical 
values of $\alpha$.   As expected, two integrable cases $\alpha =1$ and $\alpha=  \sqrt3$ have lots of degenerate energy levels while the pseudointegrable and chaotic cases have few.

With the obtained eigen-energies we can
compute $\zeta$ and $\xi$, the two parameters that we introduced
to describe quantitatively how well the two conditions (I) and (II) 
for ergodicity and mixing are satisfied in a given system. 
We first construct the two distribution functions $f(\epsilon)$ 
and $g(\Delta)$ and then compute $\zeta$ and $\xi$  using \eqref{defzeta} and \eqref{defxi}. 
The distribution function $f(\epsilon)$ together with $g(\Delta)$ 
is constructed by binning the energy levels with a width $\delta\epsilon=\hbar / T$, 
where $T$ is  the total time of a dynamical evolution. For a 
dynamical evolution of time $T$, energy levels or spacings separated by
$\delta\epsilon=\hbar / T$ can be regarded as the same. 
In our calculation, we use $T=40$ in accordance with our 
numerical study of quantum dynamics in the next section. 

The results are shown in Table.\ref{zetaxi}, where
we see clearly the values of  $\zeta$ and $\xi$ are 
strongly correlated to the classical integrability of the system. 
For integrable systems, both $\zeta$ and $\xi$ are large.  
As $\alpha$ changes and the system becomes more chaotic, 
$\zeta$ decreases almost to zero while $\xi$ is reduced by about
two orders of magnitude. These numerical results
strongly suggest that conditions (I) and (II) are largely 
satisfied by chaotic systems. 
\begin{table}
\centering
\begin{tabularx}{0.48\textwidth}{YYY}\hline
$\alpha$ & $\zeta$ & $\xi$\\ \hline
1$^*$& 0.588& 300.89\\
1.007846 & 0.032 & 14.39\\
1.015675& $0$ &8.86\\
1.077744 & $0$  &8.65\\
1.154062 & $0$  &9.25\\
1.376382 & $0$  &11.04\\
1.461725 & $0$  &11.72\\
1.662013 & 0.002 &13.33\\
1.700000 & 0.004 &13.89\\
1.718079 & 0.010 & 18.84\\
1.725067 & 0.084 &40.85\\
1.732051$^*$& 0.414& 167.03\\
\hline
\end{tabularx}
\caption{Degeneracy parameters $\zeta$ and $\xi$ for 
different values of $\alpha$. The first 1000 energy levels are used in  the calculation. 
Integrable cases are marked by superscript $*$.
Clearly, integrable cases have much larger $\zeta$ and $\xi$.}
\label{zetaxi}
\end{table}

The pseudointegrable systems are subtle. As one may have already noticed in Fig.\ref{nsd}(c)
and Table \ref{zetaxi}, the pseudointegrable case 
$\alpha=\cot{\pi/5}$ behaves very much like a chaotic system. 
However, not all pseudointegrable systems has a chaotic NSD. Some pseudointegrable triangle billiards, such as the triangle with angles $(\pi/5, 2\pi/5, 2\pi/5)$, have Possion-like 
NSDs~\cite{Miltenburg1994}. Because the peaks of their NSDs $p(s)$ are at $s=0$ which indicates large degeneracy, these 
triangle billiards should have large  $\zeta$ and $\xi$, 
and they are not ergodic or mixing. 
This difference shows that the relation
between quantum ergodicity and mixing
and classical integrability is very subtle in 
the case of  pseudointegrable systems. 

\section{Quantum Dynamical Behavior}\label{sec:db}
In this section we shall study the quantum dynamics of the
triangle billiard for a set of typical values 
of $\alpha$ to 
see whether it exhibits ergodic  or mixing behavior as described by 
Eq.(\ref{ergodic}) or Eq.(\ref{mixing}), respectively, 
and how these dynamical behaviors are dictated by 
conditions (I) and (II) via parameters $\zeta$ and $\xi$.

To study the quantum dynamical behavior, we need to  calculate the time evolution of a wave function. We use the method of  eigenstate expansion. For an arbitrary initial wave function $\psi (x,y,0)$, we expand it  in terms 
of the energy eigenstates of the Hamiltonian $\phi_k (x,y)$
\begin{equation}\label{expand}
\psi (x,y,0)=\sum_k c_k \phi_k (x,y)\,.
\end{equation}
According to the Sch\"{o}rdinger equation, the time evolution of this initial wave function is given by
\begin{equation}\label{EE}
\psi (x,y,t)=\sum_k c_k e^{\frac{-i E_k t}{\hbar}} \phi_k (x,y)\,.
\end{equation}
As the expansion coefficients can be calculated easily as
\begin{equation}
c_k=\iint\limits_{\Omega_1}\psi (x,y,0) \phi_k^* (x,y) 
\d x \d y\,,
\end{equation}
once we have obtained the expansion coefficients $c_k$, we can generate the wave function at any time.
In our study, we choose a Gaussian wave packet as 
an initial state
\begin{equation}
\psi (x,y)=\frac1{\sqrt{4\pi\sigma^2}}e^{-\frac1{4 \sigma^2} [(x-x_0)^2+(y-y_0)^2]}e^{- i2\pi(p_xx+p_yy)}\,,
\label{initial}
\end{equation}
where  $x_0=0.5, y_0=0.3, p_x=5\cos(e\pi), p_y=5\sin(e\pi)$, and $\sigma=0.02$. 
This initial state mainly occupies the first 1000 energy eigenstates.

\begin{figure}
\begin{minipage}{1.1\linewidth}
\centering
\includegraphics[width=\linewidth]{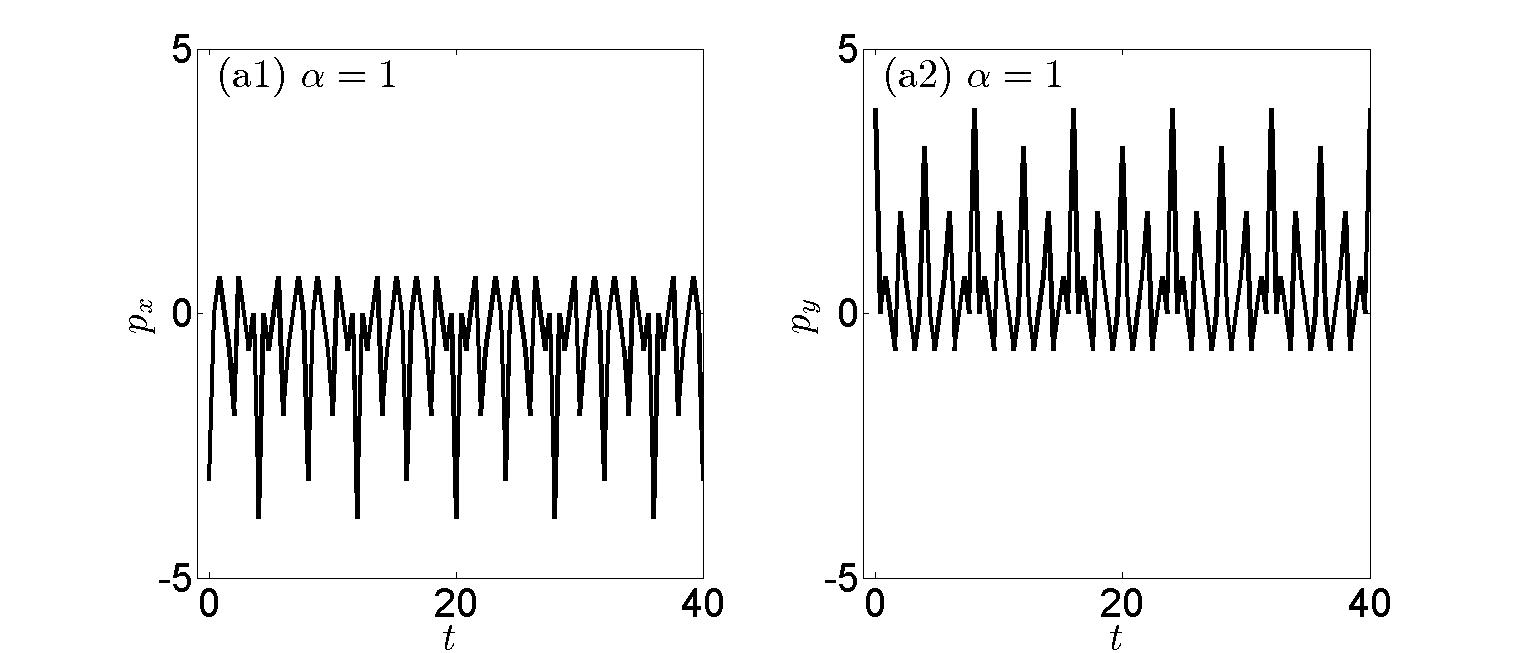}
\end{minipage}
\begin{minipage}{1.1\linewidth}
\centering
\includegraphics[width=\linewidth]{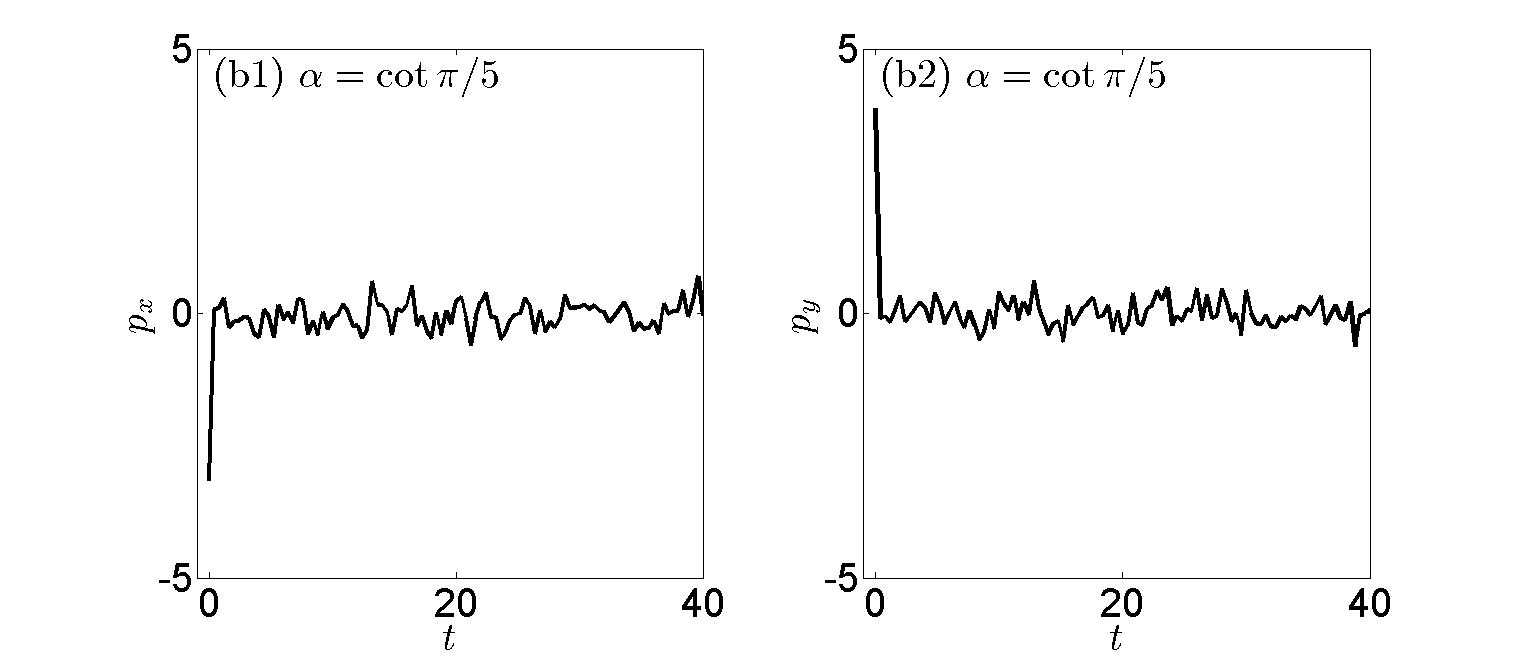}
\end{minipage}
\begin{minipage}{1.1\linewidth}
\centering
\includegraphics[width=\linewidth]{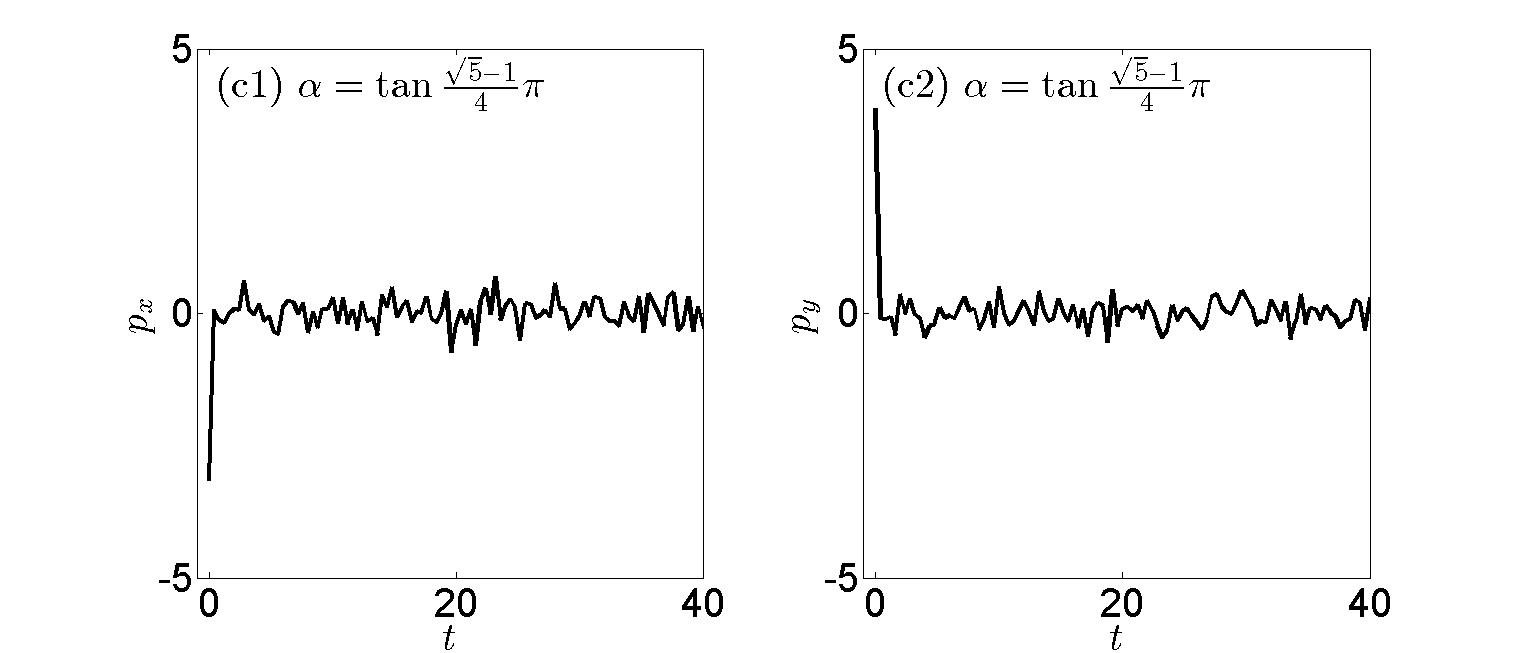}
\end{minipage}
\begin{minipage}{1.1\linewidth}
\centering
\includegraphics[width=\linewidth]{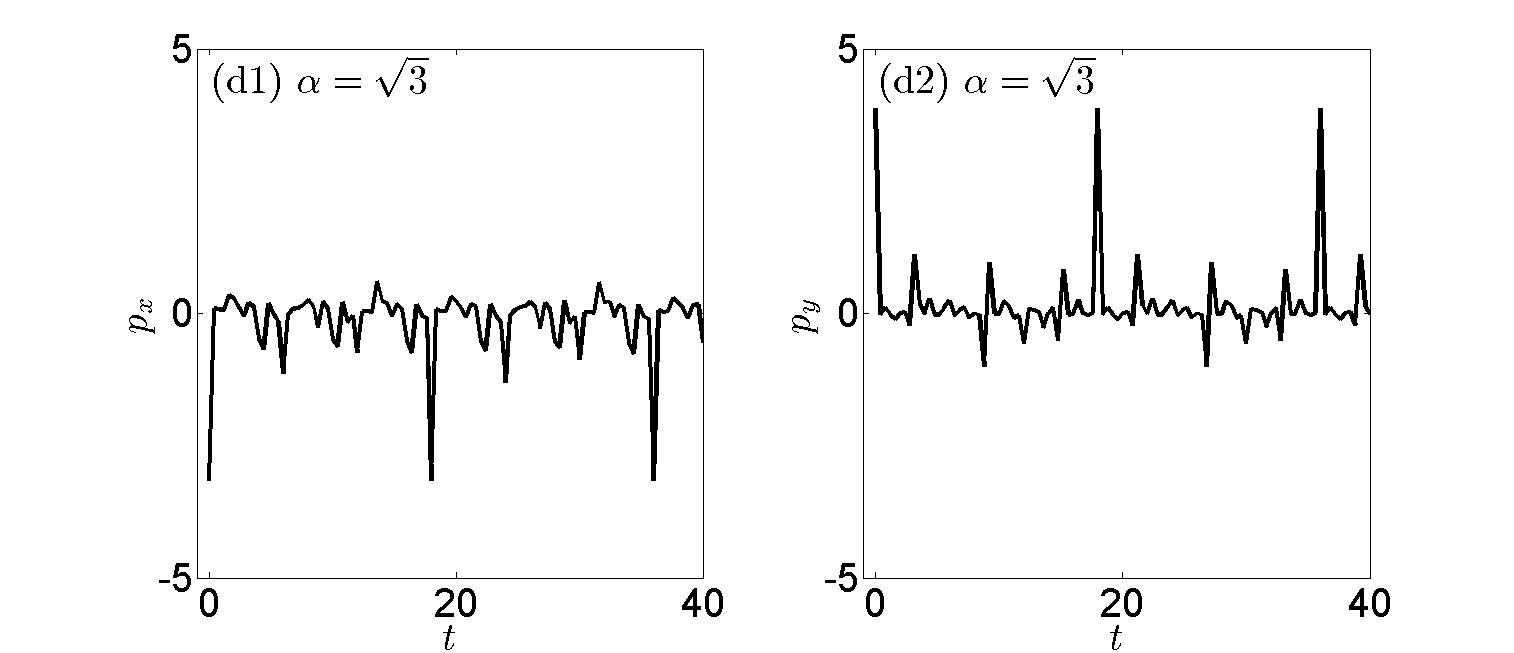}
\end{minipage}
\caption[justification=justified]{Evolution of the momentum for different $\alpha$. $\alpha=1$ and $\alpha=\sqrt{3}$ are two integrable cases.  $\alpha=\tan {\frac{\sqrt{5}-1}{4}\pi}$ is chaotic but nonergodic in classical mechanics \cite{wang2014}.  $\alpha=\cot {\pi /5}$ is pseudointegrable. }
\label{evolution}
\end{figure}

It is sufficient to focus on the momentum of the system. 
For the initial condition in Eq.(\ref{initial}) we have exactly
$\langle \vec{p} \rangle_E =0$.  The quantum dynamical 
evolutions of $\vec{p}$  are shown in Fig.\ref{evolution}
for four typical values of $\alpha$: 
$\alpha =1$, $\alpha = \sqrt{3}$, $\alpha =\cot{\pi/5}$, 
and $\alpha=\tan{\frac{\sqrt{5}-1}{4}\pi}$. 
It is clear that the evolution of $\vec{p}$ varies greatly 
with different $\alpha$. Before we discuss it in detail, 
let us first recall the classical dynamics for these four cases. 
The cases with $\alpha =1$ and $\alpha = \sqrt{3}$ 
are  integrable; $\alpha =\cot{\pi/5}$ is pseudointegrable 
and has only finite directions of $\vec{p}$ in classical dynamics which means nonergodicity. The case with $\alpha=\tan{\frac{\sqrt{5}-1}{4}\pi}$ is nonintegrable but classically nonergodic~\cite{wang2014}. 

Let us come back to the quantum dynamics in Fig.\ref{evolution}. 
For  the two integrable cases, the long-time average is apparently not equal to its microcanonical ensemble average, and the fluctuation is large as well. For other cases, the momentum  quickly relaxes to its microcanonical ensemble average and has only small fluctuations. The relaxation time is very short
and is about $\sim 10^{-1}$ for these cases. Up to $t=40$, 
we do not observe a revival or large deviation from the 
equilibrium value. These results demonstrate  that
the quantum dynamics for $\alpha=\cot{\pi/5}$ (the pseudointegrable regime) and $\alpha=\tan{\frac{\sqrt{5}-1}{4}\pi}$ are not only ergodic but also mixing. 
This is in stark contrast with their classical dynamics 
which are not even ergodic.  This  suggests
that it is easier to have quantum ergodicity and mixing 
than their classical counterparts. 

In order to check quantitatively whether a quantum system is mixing, we need to calculate the time-averaged relative  fluctuation. The averaged deviation of the momentum operator $\vec{p}$ in a given evolution time $T$ is
\begin{equation}
\langle \sigma^2_{\vec{p}} \rangle_T=\frac1T \int_0^T \left| \langle \psi (t) |\vec{p}|\psi(t) \rangle-\sum_k |c_k^2|\langle \phi_k |\vec{p}|\phi_k \rangle \right|^2\d t.
\end{equation}
Considering $E=\frac{\vec{p}^2 }{2}$, the  relative fluctuation of $\vec{p}$ is
\begin{equation}\label{AvgFluc}
F^2 =\frac{\langle \sigma^2_{\vec{p}} \rangle_T} {|\vec{p}|^2} = \frac{\langle \sigma^2_{\vec{p}} \rangle_T } {2E}.
\end{equation}
The relaxation time scale is $\sim 10^{-1}$ and the oscillating  
period for integrable systems is $\sim 10^{1}$. Considering that longer evolution time may lead to large numerical error and unreliable result, we choose $T=40$. This time length is much larger than relaxation time, and it is also long enough for us to see if there are frequent recurrences.  

\begin{figure}
\centering
\includegraphics[width=1.0 \linewidth]{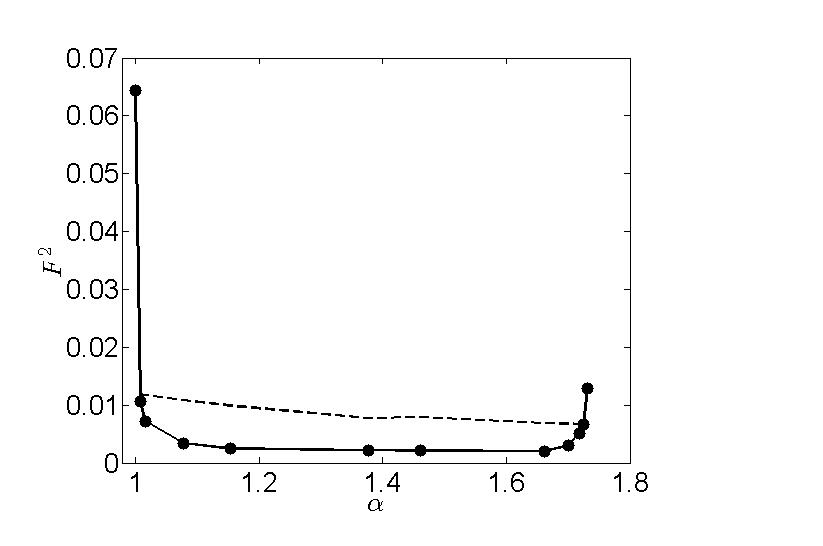}
\caption{Averaged relative fluctuations of the momentum via 
$\alpha$.  The vertical axis $F^2$ indicates the averaged relative fluctuation as defined in \eqref{mixing} and \eqref{AvgFluc}. The solid points are the numerical results; the solid line is just a guidance; the dashed line is the upper bound $\tr \rho^2_{\text{mc}} $ in \eqref{mixing}. }
\label{fluc}
\end{figure}

The results of relative fluctuation for different $\alpha$ are shown in Fig.\ref{fluc}. The dashed line is  $\tr \rho^2_{\text{mc}} $, 
the upper bound in \eqref{mixing}. 
It can be clearly seen that for $\alpha$ away from $1$ or $\sqrt{3}$, the averaged fluctuation is small, and 
the inequality \eqref{mixing} is satisfied. 
This confirms the intuitive picture in Fig. \ref{evolution}
that the quantum  dynamics is mixing. When $\alpha$ approaches the two integrable cases, $\alpha =1$ and $\alpha =\sqrt{3}$, the averaged fluctuation becomes much larger, and the inequality \eqref{mixing} is violated. In fact,  at these two integrable cases, the averaged fluctuation reaches two local maxima. There 
is a rather rapid transition from mixing to nonmixing while the system is tuned from nonintegrable to integrable. 

\begin{figure}
\centering
\includegraphics[width=1.0 \linewidth]{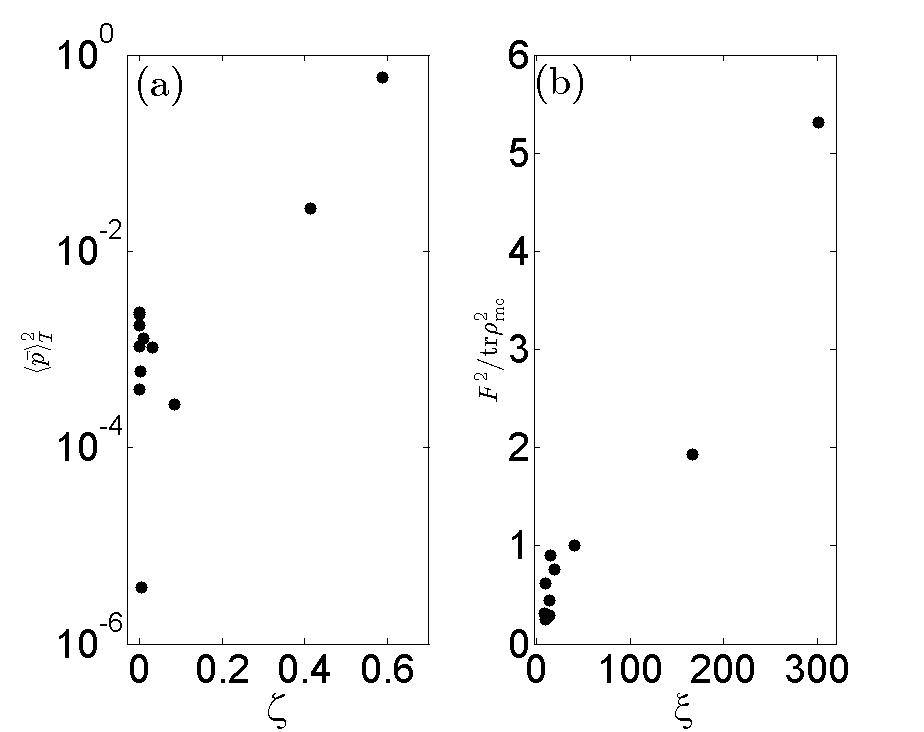}
\caption{(a)The square of time-averaged $\vec{p}$ at $T=40$ with different $\zeta$. The vertical axis  is in log scale. Note that $\langle \vec{p} \rangle _E = 0$. (b)Averaged relative fluctuation vs. $\xi$. No mixing when $F^2 / \tr \rho^2_{\text{mc}} > 1$. }
\label{xifluc}
\end{figure}

The quantum dynamic behavior shown in Fig. \ref{evolution} 
are dictated by conditions (I) and (II). This can be seen clearly
in Fig.\ref{xifluc}, where the square of time averaged momentum $\langle \vec{p} \rangle_T^2$
and its relative fluctuations are plotted against 
the two degeneracy parameters $\zeta$ and $\xi$, respectively. 
It is clear from the figure that for systems with significantly non-zero $\zeta$, the time average of the momentum significantly deviates from the microcanonical ensemble value. 
We can see a strong positive correlation between the relative 
fluctuation $F^2/\tr \rho^2_{\text{mc}}$ and $\xi$ as well. These results illustrate that 
systems with small $\zeta$ and $\xi$ have
ergodic and mixing quantum dynamics, respectively. 
Our definitions of quantum ergodicity and mixing 
with conditions (I) and (II) are legitimate.

\section{Discussion and Conclusion}\label{sec:conclusion}
Let us summarize what we have done. We have given our 
own definitions of ergodicity and mixing for quantum systems
with conditions (I) and (II).  It can be rigorously proved 
that these two conditions  lead to quantum dynamical 
behaviors which are  described by Eq.(\ref{ergodic}) and Eq.(\ref{mixing}) and 
are reminiscent of classical ergodic and mixing dynamics, respectively.  
Through an example,  the triangle billiard, we have further shown that although 
both conditions (I) and (II), which are characterized by $\zeta$ and $\xi$,
are related to classical integrability, there are differences. 
The most important is that a system whose classical dynamics is neither
ergodic nor mixing can be both ergodic and mixing in its quantum dynamics.  

\begin{figure}[th]
\begin{minipage}{0.49\linewidth}
\centering
\includegraphics[width=\linewidth]{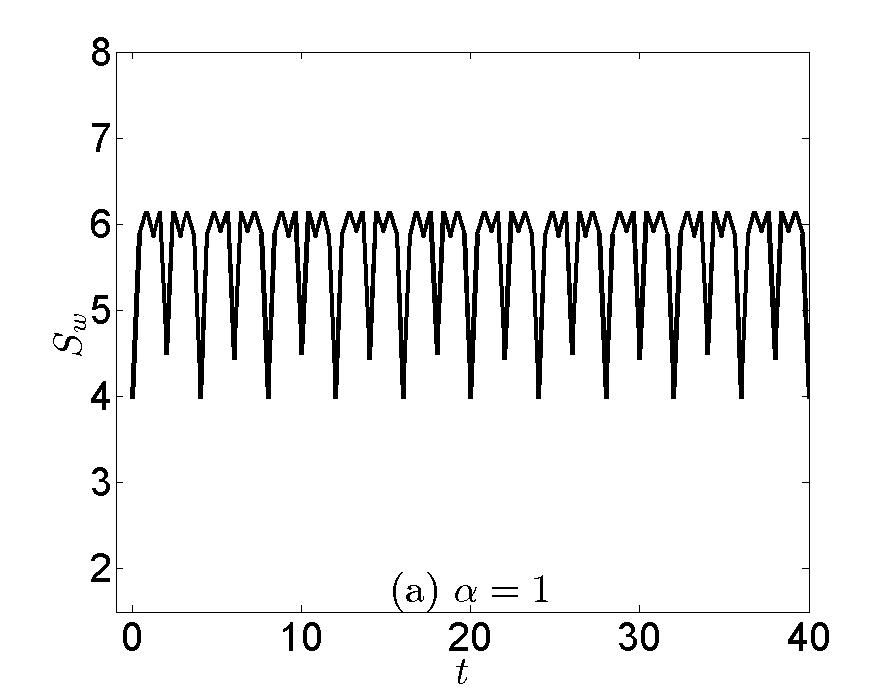}
\end{minipage}
\begin{minipage}{0.49\linewidth}
\centering
\includegraphics[width=\linewidth]{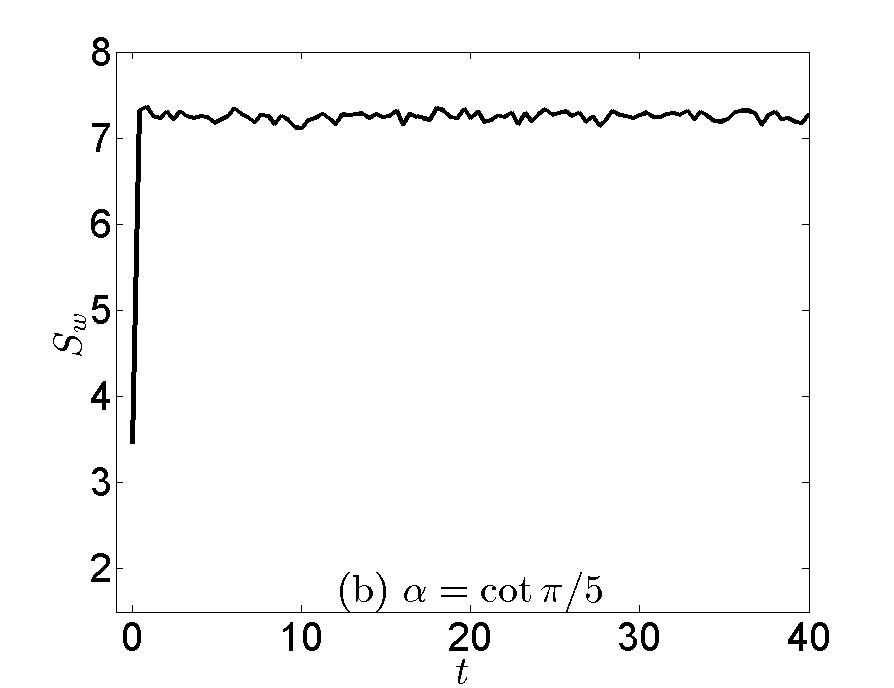}
\end{minipage}
\begin{minipage}{0.49\linewidth}
\centering
\includegraphics[width=\linewidth]{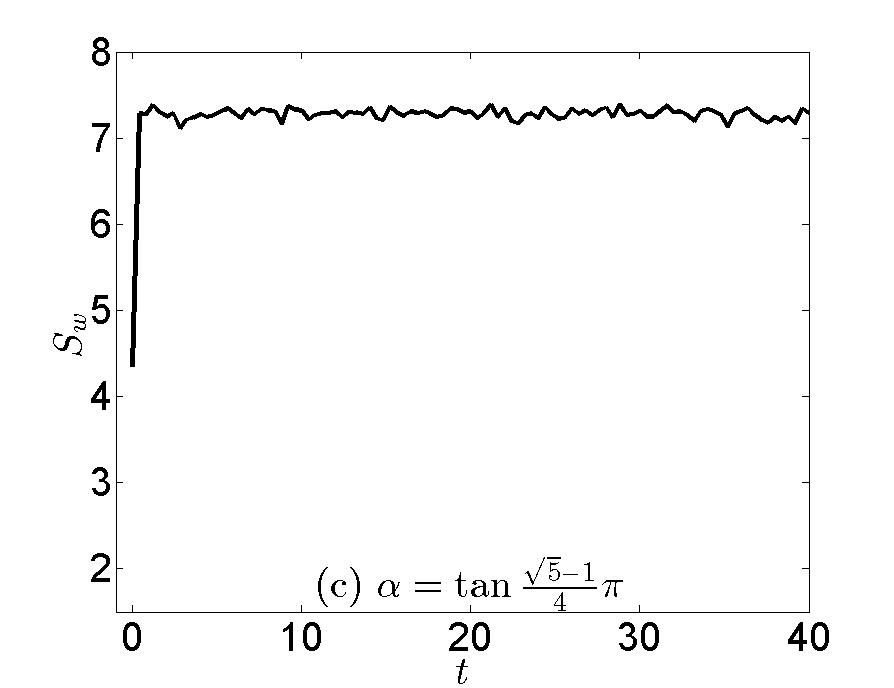}
\end{minipage}
\begin{minipage}{0.49\linewidth}
\centering
\includegraphics[width=\linewidth]{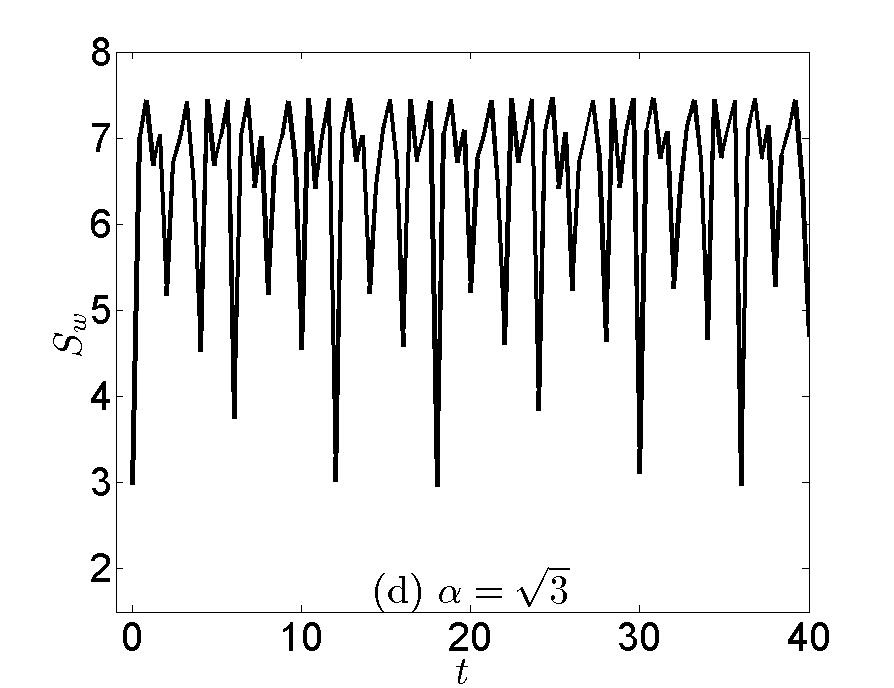}
\end{minipage}
\caption{Evolution of the entropy $S_w$ for different $\alpha$.}
\label{entropy}
\end{figure}

Classical dynamics has  an ergodic hierarchy~\cite{EH2,EH,Penrose}, which is 
\begin{equation}
{\rm Bernoulli}\subset {\rm Kolmogorov} \subset {\rm Mixing} \subset {\rm Ergodic}\,.
\end{equation}
Now mixing and ergodicity have their quantum counterparts. In particular, 
we have similar relation: quantum mixing systems are a subset of quantum 
ergodic systems.  It is possible to expand this quantum ergodic hierarchy to 
three. We define a quantum system is {\it equilibrable} if the system satisfies
\begin{multline}
E_m + E_n - E_k - E_l = E_{m'} + E_{n'} - E_{k'} - E_{l'}~{\rm and}~ \\
\{m, n\} \cap \{k, l\} = \emptyset
\Rightarrow \{m, n\} = \{m', n'\}  \{k, l\} = \{k', l'\}\,.\\
\tag{III}
\end{multline}
This condition implies that there is no degeneracy in the gaps of energy gaps. 
One can find the full implication of this condition in Ref.\cite{han2014}.  Here we briefly summarize. 
The entropy for a quantum pure state $\hat{\rho} \equiv | \psi \rangle \langle \psi |$ is defined as~\cite{han2014}
\begin{align}
S_w \equiv & - \sum_{\bm{q}_i, \bm{p}_j} \langle \psi | \bm{W}_{\bm{q}_i, \bm{p}_j}| \psi \rangle \ln \langle \psi | \bm{W}_{\bm{q}_i, \bm{p}_j}| \psi \rangle \notag \\
\equiv & - \sum_{\bm{q}_i, \bm{p}_j}\tr (\hat{\rho} \bm{W}_{\bm{q}_i, \bm{p}_j}) \ln\tr (\hat{\rho} \bm{W}_{\bm{q}_i, \bm{p}_j}), 
\end{align}
where $\bm{W}_{\bm{q}_i, \bm{p}_j} \equiv |w_{\bm{q}_i, \bm{p}_j} \rangle \langle w_{\bm{q}_i, \bm{p}_j}|$ is the projection onto Planck cells in quantum phase space at position $\bm{q}_i$ and momentum $\bm{p}_j$ and $\{|w_{\bm{q}_i, \bm{p}_j} \rangle\}$ is a complete set of 
Wannier functions. This entropy $S_w$ will change with time. An inequality regarding 
the relative fluctuation of entropy $S_w$, similar to Eq.(\ref{mixing}), 
was proved in Ref.\cite{han2014} with condition (III). This inequality means that
a quantum system with small entropy $S_w$ will relax dynamically to a state whose
entropy $S_w$ is maximized and stay at this maximized value with small fluctuations. 
This is illustrated with the triangle billiard in  Fig.\ref{entropy}. 
In this figure we 
see that the entropy $S_w$ of the integrable cases, for which condition (III) is not satisfied, 
fluctuates periodically with large amplitude  and does not stay at the maximum value. For cases with $\alpha=\tan{\frac{\sqrt{5}-1}{4}\pi}$ and $\alpha =\cot{\pi/5}$, where condition (III) is largely 
satisfied, the entropy quickly relaxes to the maximum value and stays there with small fluctuations. 
These results demonstrate that a quantum system that satisfies condition (III) is capable of
equilibrating to a state where not only its observables fluctuate around its equilibrium 
value with small amplitude  but also its entropy is maximized. This is the reason 
that we call such a quantum system {\it equilibrable}. In this way we have a quantum 
ergodic hierarchy
\begin{equation}
{\rm {\it Equilibrable}} \subset {\rm Mixing} \subset {\rm Ergodic}\,.
\end{equation}
We do not call it quantum  Kolmogorov as we do not see an apparent connection
to the classical Kolmogorov mixing system at this moment.  It would be very interesting
to find a quantum system which is equilibrable but not mixing. 

Finally we conclude with expectations of more future work to follow. 
We have given precise definitions of quantum ergodicity and mixing which
are in accordance with our usual understanding of ergodicity and mixing. 
We have illustrated with single-particle billiard systems. It would be very 
interesting to further examine them in true many-body quantum 
systems~\cite{Presilla1996,Benet2010}, where the thermodynamics limit can be
considered. 

\section{Acknowledgments}
We thank Zongping Gong for helpful discussion. 
This work is supported by the National Basic Research Program of China (Grants No. 2013CB921903 and No. 2012CB921300) and the National Natural Science Foundation of China (Grants No. 11274024, No. 11334001, and No. 11429402)..

\appendix
\section{Calculation of eigen-energies and eigenstates in the triangle billiards}\label{app}
We use the exact diagonalization method to calculate the eigen-energies and 
eigenstates: ({\it i}) choose an appropriate set of basis;  ({\it ii}) calculate the Hamiltonian matrix elements in the basis; ({\it iii}) numerical diagonalization that results the eigenenergies and eigenstates.

In order to reduce the numerical error to an acceptable range, 
we choose the basis as follows
\begin{equation}
|m,n\rangle=\frac2{l \sqrt{\alpha}}(\sin\frac{m\pi x}{l}\sin\frac{n\pi y}{\alpha l} - \sin\frac{n\pi x}{l}\sin\frac{m\pi y}{\alpha l})\,,
\end{equation}
This choice is similar to that in Ref.\cite{Miltenburg1994}. This basis is complete and orthogonal.  It is easy to check that all these base functions $\ket{m,n}$ are zero on the boundaries 
of the triangle. The elements of the Hamiltonian matrix can  be computed analytically
\begin{widetext}
\begin{eqnarray}
\langle m_1,n_1 |\hat{H} |m_2,n_2 \rangle &=&\frac{h^2}{2ml^2} \Big\{ \left(m_2^2+\frac{n_2^2}{\alpha^2} \right) \big[ I(m_1,n_1,m_2,n_2)-I(n_1,m_1,m_2,n_2)\big] \nonumber\\
&& -\left(n_2^2+\frac{m_2^2}{\alpha^2} \right) \big[ I(m_1,n_1,n_2,m_2)-
I(n_1,m_1,n_2,m_2)\big] \Big\}, 
\end{eqnarray}

where
\begin{equation}\label{int:sin}
\begin{aligned}
&I(m,n,p,q) = \int_0^1 \d x \int_0^x \d y \sin(m\pi x) \sin(n\pi y) \sin(p\pi x) \sin(q \pi y)   \\
&= \left\{ \begin{aligned}
&\frac18,\hspace{39em} \text{if }m=p ~\&~ n=q  \\
&\frac{1}{8\pi^2}(1-(-1)^{m+n+p+q})\left[ -\frac{\{p-m+n+q\}^{-1}+\{q+n-p+m\}^{-1}-\{p+m+n+q\}^{-1}-\{n+q-p-m\}^{-1}}{q+n} \right. \\
&\left. 
+ \frac{2}{(m+p)^2} -\frac2{(m-p)^2} \right],\hspace{28.61em} \text{if }m \neq p ~\&~ n=q \\
&\frac{1}{8\pi^2}(1-(-1)^{m+n+p+q})\left[ -\frac{\{p-m+n+q\}^{-1}+\{q+n-p+m\}^{-1}-\{p+m+n+q\}^{-1}-\{n+q-p-m\}^{-1}}{q+n}
 \right. \\
&\left.  + \frac{\{p-m+q-n\}^{-1}+\{q-n-p+m\}^{-1}-\{m+p+q-n\}^{-1} -\{q-m-n-p\}^{-1}}{q-n}
 \right],
\hspace{1.255em} \text{if }m \neq p ~\&~ n \neq q
\end{aligned}\right.
\end{aligned}
\end{equation}
\end{widetext}
with the curly braces $\{ \cdot\}^{-1}$ representing
\begin{equation}
\{ z \}^{-1}=\left\{
\begin{aligned}
&0, &\text{if }z=0,  \\
&\frac1z, &\text{if }z \neq 0.  \\
\end{aligned}\right.
\end{equation}
This notation is used only in \eqref{int:sin} to simplify the expression.

After the above derivation, we take a cutoff in $n_1,n_2,m_1,m_2$ and choose $h=m= l=1$ to calculate the elements of Hamiltonian matrix.  
The eigenenergies and eigenstates can be obtained after diagonalization of the Hamiltonian matrix. 
As the elements of Hamiltonian matrix are explicit, the error of the eigenenergies and eigenstates mainly arises from the cutoff of $n_1,n_2,m_1,m_2$. In our calculation, the number of basis is set as 8500. Changing this number from 6000 to 10000 only cause a $\sim 0.01\%$ relative variation of eigenenergies  (in the unit of $h^2/ml^2$). This indicates that the error in the numerical results of eigenenergies is around $0.01\%$, which is accurate enough for our analysis.  

\section{Proofs of  Long-time Quantum Ergodic and Mixing Behaviors, \eqref{ergodic} and \eqref{mixing}}
\label{appProve}
In this appendix we provide the proofs of \eqref{ergodic} and \eqref{mixing}, which
concern the long-time ergodic and mixing behavior in  quantum  systems, respectively.  
The original versions of the proofs can be found 
in Ref.\cite{neumann1929,peres1984,reimann2008,short2011}.

Consider a quantum system that starts with the following initial condition,
\begin{equation}
\ket{\psi(0)}=\sum_n c_n \ket{\phi_n}\,,
\end{equation}
where $\ket{\phi_n}$'s are the energy eigenstates. 
At time $t$, the wave function becomes
\begin{equation}
\ket{\psi(t)}=\sum_n c_n e^{-iE_nt/\hbar}\ket{\phi_n}\,.
\end{equation}
The corresponding density matrix is then 
\begin{eqnarray}
&&\hat{\rho}(t) =\ket{\psi(t)}\bra{\psi(t)}\nonumber\\
&=&\sum_{m,n}\rho_{nm} e^{-i(E_n-E_m)t/\hbar}\ket{\phi_n}\bra{\phi_m}\,.
\end{eqnarray}
where $\rho_{nm}=c_m^* c_n$.  In an ergodic system where condition (\ref{nde}) is satisfied,
we have
\begin{eqnarray}
\langle \hat{\rho}(t) \rangle_T=&&\left\langle \sum_{m,n}\rho_{nm}e^{-i(E_n-E_m)t/\hbar}|\phi_n\rangle \langle \phi_m|\right\rangle_T \nonumber\\
= &&\sum_{m,n}\rho_{nm}\langle e^{-i(E_n-E_m)t/\hbar} \rangle_T|\phi_n\rangle \langle \phi_m|\nonumber\\
=&&\sum_{m,n}\rho_{nm}\delta_{E_n,E_m}|\phi_n\rangle \langle \phi_m|\nonumber\\
=&&\sum_{m,n}\rho_{nm}\delta_{n,m}|\phi_n\rangle \langle \phi_m|\nonumber\\
=&&\sum_m \rho_{mm} |\phi_m \rangle \langle \phi_m|\nonumber\\
=&&\hat{\rho}_{\text{mc}}, 
\end{eqnarray}
which is exactly the micro-canonical ensemble that we introduced in \eqref{rhomc}. 
Therefore, for an observable $\hat{A}$, 
\begin{equation}
\langle \hat{A} \rangle_T=\langle\tr \hat{A} \hat{\rho}(t)\rangle_T=\tr [\hat{A}\langle \hat{\rho}(t) \rangle_T]=\tr \hat{A}\hat{\rho}_{\text{mc}}=\langle \hat{A} \rangle_E\,. 
\end{equation}
This is the proof of \eqref{ergodic}.

We now compute the standard deviation of $\hat{A}$.  
\begin{eqnarray}
&&\langle\sigma_A^2 \rangle_T
=\left\langle \Big| \langle \hat{A}(t)\rangle- \langle \hat{A}\rangle_E\Big|^2 \right\rangle_T \nonumber\\
&&=\left\langle \left|\langle \hat{A} (t) \rangle\right|^2 \right\rangle_T -\left|\langle \hat{A} \rangle_E\right|^2\nonumber\\
&&=\sum_{k\neq l, m \neq n }\rho_{lk}^*\rho_{nm} \left\langle e^{-i[(E_n-E_m)-(E_l-E_k)]/\hbar }\right\rangle_T A_{mn} A_{kl}^*\nonumber \\
\end{eqnarray}
When the quantum system is mixing, that is, both conditions  (\ref{nde}) and (\ref{ndg}) are satisfied,  we have 
\begin{eqnarray}
&\langle\sigma_A^2 \rangle_T&=\sum_{k\neq l, m \neq n }\rho_{kl}\rho_{nm}  \delta_{E_n-E_m,E_l-E_k}  A_{mn}A_{lk}  \nonumber\\
&&=\sum_{k\neq l, m \neq n }\rho_{kl}\rho_{nm}  \delta_{mk}\delta_{nl} A_{mn}A_{lk} \nonumber \\
&&=\sum_{m\neq n} \rho_{mn}\rho_{nm} A_{mn}A_{nm}\nonumber\\
&&=\sum_{m,n} |c_m|^2 |c_n|^2 A_{mn}A_{nm}- \sum_{n}|c_n|^4 |A_{nn}|^2\nonumber\\
&&\leq\sum_{m,n} \rho_{mm}\rho_{nn} A_{mn}A_{nm}\nonumber\\
 &&= \tr\hat{A}\hat{\rho}_{\text{mc}}\hat{A}^\dag\hat{\rho}_{\text{mc}}=\tr\left[(\hat{\rho}_{\text{mc}}\hat{A}^\dag)^\dag(\hat{A}^\dag\hat{\rho}_{\text{mc}})\right]\,,  \nonumber\\
\end{eqnarray}
where we have used $A_{mn}=\langle \phi_m |\hat{A}|\phi_n \rangle$.  We define a scalar product for two operators $\hat{P}, \hat{Q}$ as $\tr (\hat{P}^\dag \hat{Q})$.  Using 
the Cauchy-Schwartz inequality for operators with such scalar product, 
we have~\cite{short2011}
\begin{eqnarray}
&&\langle\sigma_A^2  \rangle_T\leq \sqrt{\tr\left[(\hat{\rho}_{\text{mc}}\hat{A}^\dag)^\dag(\hat{\rho}_{\text{mc}}\hat{A}^\dag)\right]\tr\left[(\hat{A}^\dag\hat{\rho}_{\text{mc}})^\dag(\hat{A}^\dag\hat{\rho}_{\text{mc}})\right]}
\nonumber\\
&&=\sqrt{\tr(\hat{A}^\dag\hat{A}\hat{\rho}_{\text{mc}}^2)\tr(\hat{A}\hat{A}^\dag\hat{\rho}_{\text{mc}}^2)} \nonumber\\
&&\leq \|\hat{A}\|^2 \tr \hat{\rho}_{\text{mc}}^2, 
\end{eqnarray}
where $\| \hat{A} \|^2=\sup \{\langle \psi | \hat{A}^\dag \hat{A} | \psi \rangle : \ket{\psi} \in \mathscr{H}\}$ is the upper limit of
the expectation value of $\hat{A}^2$ in the Hilbert space. 
Finally, we have for the fluctuation
\begin{equation}
F^2_A \equiv \frac{\langle\sigma_A^2  \rangle_T}{\|\hat{A}\|^2} = \frac{\left\langle \Big| \langle \hat{A}(t)\rangle- \langle \hat{A}\rangle_E\Big|^2 \right\rangle_T}{\|\hat{A}\|^2}
 \leq \tr \hat{\rho}^2_{\text{mc}}\,.
\end{equation}
This is the proof of \eqref{mixing}.

\bibliography{Summary-4}

\end{document}